\documentclass[aps,prl,floats,twocolumn,superscriptaddress,amssymb,amsmath,showpacs]{revtex4}
\usepackage{psfrag}
\usepackage{graphicx}
\usepackage{bm}
\begin{document}
\title{Fluctuations of Fluctuation--Induced  ``Casimir'' Forces}
\author{Denis Bartolo}
\address{Laboratoire de Physico-Chimie Th\'eorique, UMR CNRS 7083,
ESPCI, 10 rue Vauquelin, F-75231 Paris Cedex 05, France}
\author{Armand Ajdari}
\address{Laboratoire de Physico-Chimie Th\'eorique, UMR CNRS 7083,
ESPCI, 10 rue Vauquelin, F-75231 Paris Cedex 05, France}
\author{Jean-Baptiste Fournier}
\address{Laboratoire de Physico-Chimie Th\'eorique, UMR CNRS 7083,
ESPCI, 10 rue Vauquelin, F-75231 Paris Cedex 05, France}
\address{F\'ed\'eration de recherche FR CNRS 2438 ``Mati\`ere et Syst\`emes
complexes''}
\author{Ramin Golestanian}
\address{Institute for Advanced Studies in Basic Sciences, Zanjan
45195-159, Iran}
\address{Institute for Studies in Theoretical Physics and
Mathematics, P.O. Box 19395-5531, Tehran, Iran}

\begin{abstract}
The force experienced by objects embedded in a correlated medium
undergoing thermal fluctuations---the so-called
fluctuation--induced force---is actually itself a fluctuating quantity. 
We compute the corresponding probability distribution and
 show that it is a Gaussian centered on
the well-known Casimir force, with a non-universal
standard deviation that can be typically as large as the mean force
itself. The relevance of these results to the experimental
measurement of fluctuation-induced forces is discussed,
as well as the influence of the finite temporal resolution
of the measuring apparatus.
\end{abstract}
\pacs{05.40.-a, 68.60.Dv, 82.70.-y}

\maketitle
\newcommand{\kt}{k_\mathrm{B}T}
\newcommand{\D}{{\cal D}}
\newcommand{\E}{{\cal E}}
\newcommand{\Z}{{\cal Z}}
\newcommand{\p}{{\cal P}}
\newcommand{\T}{{\cal T}}
 \newcommand{\h}{{\cal H}}
\newcommand{\F}{{\cal F}}
\newcommand{\Tmes}[1]{{\overline{{\cal  T}^{#1}}}}
\newcommand{\tqmes}[1]{{\overline{\tau_{\bf q}}^{\,#1}}}
\newcommand{\deron}[2]{\frac{\partial #1}{\partial #2}}
\newcommand{\limit}[2]{\xrightarrow[#1\rightarrow #2]{}}
\newcommand{\integrale}[4]{\int_{#1}^{#2}\!d^{#3}#4\,}
\newcommand{\integraleq}[4]{\int_{#1}^{#2}\!\frac{d^{#3}#4}{(2\pi)^{#3}}\,}
In 1948, Casimir predicted that two uncharged conducting plates
facing each other in vacuum are subject to a long-range universal
attraction~\cite{Casimir}. The latter is due to the modification of
the quantum electromagnetic fluctuation spectrum due to the
boundary conditions imposed by the conducting plates. In the
last fifteen years, the concept of  Casimir
force has been extended to thermally excited elastic forces between objects
embedded in a medium with scale-free fluctuations altered by the
presence of the objects~\cite{raminrmp}. Soft matter examples comprise
inclusions in complex fluids undergoing thermal fluctuations~\cite{raminrmp},
e.g. fluid membranes~\cite{goulian,pauljbf} and critical 
mixtures~\cite{dietrich}, or interfaces bounding complex fluids, e.g., liquid
crystals~\cite{aathese,ziherl} and superfluids~\cite{kardarli, krech}.  One of
the main features of these Casimir interactions is their universality. In a
given geometry and at a given temperature, the fluctuation--induced forces
depend only on the universality class of the fluctuating medium and on the
nature of the imposed boundary conditions (the material's elastic constants and
the coupling strength of the boundaries are irrelevant).

While the development of micro- and nanoscale experiments has
allowed precise direct verifications of the original vacuum's
Casimir force~\cite{lamoreaux,mohideen}, experimental characterization
of thermal fluctuation--induced forces in soft-matter has been
scarce~\cite{garcia,law}. The main difficulty with measuring these forces is usually
imputed to the presence of stronger background elastic and van der Waals forces.
Another important point which has been
overlooked, is that the fluctuation-induced force is itself a fluctuating
quantity: what is commonly referred to as the ``Casimir force'' is only its
\textit{ensemble average}.
The fluctuating nature of this force should be taken into proper account before
their complete characterization can be achieved. Similar
consideration hold for the experimental measurement of
quantum Casimir forces~\cite{barton,lamoreaux,mohideen}, but we
focus here only on the classical regime.

In this Letter we fully characterize the fluctuations of these thermally
excited elastic forces.
Generically, we consider in a space of dimension~$d$, a
fluctuating medium described by a scalar field~$\phi$ with an elastic energy
density proportional to $(\vec\nabla\phi)^2$.  For instance, $\phi$ could
represent the deviation of the director of a nematic liquid
crystal~\cite{deGennes} from a uniform orientation (in the
one-elastic-constant approximation)~\cite{aathese} or the
composition of a mixture at its critical
point~\cite{krech}.
\begin{figure}
\includegraphics[width=6cm]{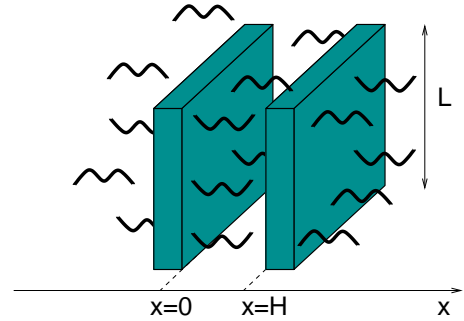}
\caption{\label{geometry} Two identical macroscopic plates normal
to the $x$ axis immersed in a fluctuating correlated medium.}
\end{figure}
We consider two plates of lateral
extension $L$ separated by a distance $H$ (see
Fig.~\ref{geometry}) that impose $\phi=0$ on their boundary.  We
show that the fluctuating force is characterized by a Gaussian
distribution for $d>1$ centered on the
usual Casimir force $\langle
F\rangle \sim k_B T L^{d-1}/H^d$.
We characterize the corresponding dispersion by the ``noise-over-signal ratio,''
i.e. the width $\Delta F$ of the distribution over the mean $\langle
F\rangle$. By calculating $\Delta F$ we show that this ratio scales as
\begin{equation}
\frac{\Delta F}{\langle F\rangle}\sim
\left(\frac{H}{L}\right)^{\frac{d-1}{2}}
\left(\frac{H}{a}\right)^{\frac{d+1}{2}}.\label{primeresult}
\end{equation}
In the above result $a$ corresponds to a microscopic length scale
at which either the continuum description breaks down or the boundary
conditions are not efficiently set by the plates~\cite{plasma}.
We also consider the dynamics of the field and the role of the
finite temporal resolution $\tau$ of a measurement apparatus. If
$\tau_{a}$ is the microscopic time-scale associated with the
microscopic length $a$, we show that the noise to signal ratio is
significantly reduced if the apparatus' resolution induces some
additional averaging  (i.e. $\tau > \tau_a$ ). More precisely,
the average is still $\langle F\rangle$, but the noise to signal
ratio now scales as: \begin{equation} \frac{\Delta
F_\tau}{\langle F_\tau\rangle}\sim
\left(\frac{H}{L}\right)^{\frac{d-1}{2}}
\left(\frac{H}{a}\right)^{\frac{d+1}{2}}
\sqrt{\frac{\tau_{a}}{\tau}}.\label{2emeresult} \end{equation}
The relevance of the above results to experimental situations is
discussed at the end of the paper.

We consider an elastic Hamiltonian of the form
\begin{equation}
\label{energie} \h[\phi]=\frac{K}{2}\!\int\!d^d\vec R\left[
\vec\nabla\phi(\vec R)\right]^{2},
\end{equation}
with $\vec{R}=(x,\mathbf{r})$, and $x$ the coordinate in the
direction normal to the plates. We assume that the plates impose
Dirichlet boundary conditions, i.e.,
$\phi(x=0,\mathbf{r})=\phi(x=H,\mathbf{r})=0$. The whole system is
in contact with a heat bath imposing the temperature $T$. For this
minimal model, the free energy $\F(H)$ of the system can be
exactly calculated as a function of the interplate distance
$H$~\cite{kardarli}. The Casimir force $\langle
F(H)\rangle\equiv-\partial \F /\partial H$, i.e. the ensemble
average of the fluctuation-induced force, is given by the
universal formula
\begin{equation}
\langle F \rangle=-\kt \; A_d \;{L^{d-1} \over
H^{d}},\label{moyenneF}
\end{equation}
where the prefactor $A_d=\Gamma(d/2)\zeta(d)/(4\pi)^{d/2}$ depends only on the
spatial dimension $d$.

\begin{figure} \includegraphics[height=2.5cm]{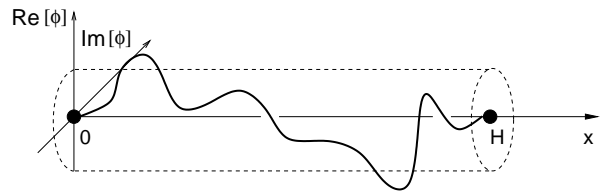}
\caption{Typical configuration of a fluctuating string pinned at
$x=0$ and $x=H$. The dashed cylinder sketches the effective {\em
cage} imposed by the confining quadratic potential of curvature
$Kq^{2}/L^{d-1}$.} \label{figure2}
\end{figure}

We shall pursue another route to calculate the Casimir
interaction, which yields a clear intuitive picture of the origin
of the fluctuation--induced force, and permits evaluation of its
fluctuations. We use the stress tensor
associated with the field \cite{landau}:
$T_{ij}=\E\delta_{ij}-(\partial_{i}\phi)
[\partial\E/\partial(\partial_{j}\phi)]$, where
$\E=\frac{1}{2}K(\vec\nabla\phi)^2$ is the energy density
associated with Eq.~(\ref{energie}). We first restrict our
attention to the effect of the interplate medium
on the plate located at $x=H$.
For a given conformation of the field, the projection
$F^<(H)$ on the $x$-axis of the elastic force exerted on the
inner side of this plate is given by the
integral of $T_{xx}$, which reduces to
\begin{equation}
F^<(H)=\frac{K}{2}\!\int\!d^{d-1}{\bf r}\left
[\partial_x\phi\left(H,{\bf r}\right)\right]^2,
\end{equation}
since $\phi$ vanishes all along the plate. Note that this quantity
is always positive, therefore the field always pushes on the
plate.  Fourier transforming the field in the transverse
direction: $\phi_\mathbf{q}(x)=\int\!d^{d-1}{\bf r}\,\phi(x,{\bf
r})\,e^{-i\mathbf{q}\cdot\mathbf{r}}$ yields
$\mathcal{H}=\sum_\mathbf{q}E_\mathbf{q}$ and
$F^<(H)=\sum_\mathbf{q} f_\mathbf{q}(H)$ with
$E_\mathbf{q}=\frac{1}{2}L^{1-d}K\int\!dx\,
[\phi_\mathbf{q}(x)(-\partial_x^2+q^2)\phi_{-\mathbf{q}}]$ and
$f_\mathbf{q}(H)=\frac{1}{2}L^{1-d}K|\partial_x\phi_\mathbf{q}(H)|^2$. These expressions
suggest an interesting picture: each $E_{\bf q}$ can be understood
as the energy of a three-dimensional string parameterized by
$[x,\mathrm{Re}(\phi_\mathbf{q}),\mathrm{Im}(\phi_\mathbf{q})]$;
each of these virtual strings has a line tension proportional to
the rigidity $K$ of the medium and is confined around the
$x$-axis by a harmonic potential of stiffness proportional to $q^2$. These
lines are pinned at $x=0$ and $x=H$ as a consequence of the
Dirichlet boundary conditions (see Fig.~\ref{figure2}) and the
force they exert on the plate located at $x=H$ is precisely
$f_\mathbf{q}(H)$. The total partition function
of the
medium between the plates is the product
$\Z=\prod_{|\mathbf{q}|<1/a}\Z_{\bf q}$ of those of the independent strings
\begin{equation}
\label{Zq}
\mathcal{Z}_\mathbf{q}=
\int_{\phi_\mathbf{q}(0)=0}^{\phi_\mathbf{q}(H)=0}
\!\!\!\!\!\!\!\D\phi_\mathbf{q}(x)\,e^{-E_{\bf q}[\phi_\mathbf{q}]/\kt},
\end{equation}
where $a^{-1}$ is the high wavevector cutoff. The
number of non-interacting strings describing our system is
from simple mode counting:
\begin{equation}
N_a\sim\left(\frac{L}{a}\right)^{d-1}. \label{Na}
\end{equation}

Determining the distribution of the elementary forces
$f_\mathbf{q}(H)$ is straightforward. Since the free energy
associated with the string fields $\phi_\mathbf{q}$ is quadratic,
the distribution of $\partial_x\phi_\mathbf{q}(H)$ is a Gaussian
with zero mean and variance $\partial_x\partial_yG_{\bf q}(H,H)$,
where $G_{\bf q}(x,y)\equiv\langle \phi_{\bf q}(x)\phi_{-\bf
q}(y)\rangle$ is the two-point correlation function~\cite{JZJ}:
\begin{equation}
G_{\bf q}(x,y)=L^{d-1}\frac{\kt}{K}
\frac{\sinh\left[q(H-u)\right]\sinh(qv)}{q\sinh(qH)},\label{green}
\end{equation}
where $u={\rm max}(x,y)$ and $v={\rm min}(x,y)$. Since
the distribution of $\partial_x\phi_\mathbf{q}$ is Gaussian, the
fluctuating elementary forces follow a $\chi^2$-distribution:
\begin{equation}
\p\left[\frac{f_\mathbf{q}(H)}{\langle
f_\mathbf{q}(H)\rangle}=f\right] = \frac{\theta(f)}{\sqrt{2\pi
f}}\exp\left(-\frac{f}{2}\right). \label{distributionfq}
\end{equation}
where $\theta$ is the Heaviside step function, and
\begin{equation}
\langle f_{ \bf q}(H)\rangle= \frac{\kt}{2}\left[{2
\over \pi a}-q\coth(qH)\right], \label{fqmoyen}
\end{equation}
The cutoff $a^{-1}$ appears in Eq.~(\ref{fqmoyen}) because of the
discontinuity of the derivative of $G_{\bf q}(x,y)$~\cite{delta0},
as a manifestation of the importance of the short wavelength
fluctuations. From Eq.~\ref{green} follows that the elementary forces
undergo large fluctuations: their standard deviation, given by
\begin{equation}
\label{fqvariance}
\Delta f_\mathbf{q}(H)=\sqrt{2}\,\langle f_\mathbf{q}(H)\rangle,
\end{equation}
compares with their average value.

We can now consider the whole space  consisting of three
independent subsystems delimited by the plates. The instantaneous
force $F(H)$ experienced by the plate at $x=H$ is the difference between two
independent stochastic variables $F^<(H)-F^<(\infty)$, where $F^<(\infty)$
obviously describes the contribution of the outer medium $x>H$.
Within the string picture
$F(H)=\sum_\mathbf{q}[f_\mathbf{q}(H)-f_\mathbf{q}(\infty)]$ thus appears as
the sum of $N_a$ independent variables.

\textit{Mean Casimir force}: From Eq.~(\ref{fqmoyen}),
the mean force $\langle F\rangle$ experienced by each plate is given by
(replacing the discrete sum by an integral and disregarding the
cutoff as the integral converges): $ \langle
F\rangle=\frac{1}{2}\kt(L/2\pi)^{d-1}
\int\!d^{d-1}\mathbf{q}
\left[q-q\coth(qH)\right]$, which yields the result quoted in Eq.
(\ref{moyenneF}) above. We thus recover the usual expression of
the Casimir force, classically obtained by differentiation of the free energy.

The cancellation of the non-universal cutoff contribution in the
former sum is easily understood in terms of the strings picture:
since a string with label $\mathbf{q}$ has a correlation length
$\simeq q^{-1}$, the strings with $q\gg H^{-1}$ do not feel the
presence of the second plate and their average contributions on
both sides of the plate cancel exactly.  Conversely,
the force imbalance
for each string with $q<1/H$ scales as $-\kt/H$, and there are
$(L/H)^{d-1}$ such strings, the product yielding the scaling
form of the Casimir
force in Eq. (\ref{moyenneF}).

\textit{Distribution of the fluctuation-induced force}: The
variance of the force, $(\Delta F)^2\equiv\langle
F^2\rangle-\langle F\rangle^2$, is the sum of the $2N_a$
elementary contributions $(\Delta f_{\bf q})^{2}$ from both
sides of the plate:
\begin{eqnarray}
\label{Fvariance}
&&(\Delta F)^2=(\kt)^2L^{d-1}\times\nonumber\\
&&\frac{1}{2}\int\!\frac{d^{d-1}\mathbf{q}}{(2\pi)^{d-1}} \left\{ \left[{2
\over \pi a}-q\coth\left(qH\right)\right]^2+\left({2 \over \pi
a}-q\right)^2\right\}, \qquad
\end{eqnarray}
where the integral must be limited to $|\mathbf{q}|<a^{-1}$
to be defined. When $H\gg a$, the leading behavior is given
by the limit $H\to\infty$, i.e., $(\Delta F)^2\sim (\kt)^2
L^{d-1}\int_{q<1/a} [(\pi a)^{-1}-q]^2 q^{d-2} d q$, which
yields
\begin{equation}
\label{varianceF}
\Delta F\sim\sqrt{N_a}\,\frac{\kt}{a},
\end{equation}
where again $N_a$ is the number of transverse Fourier modes
defined in Eq. (\ref{Na}) above. Contrary to the Casimir (average)
force, the variance of the fluctuation--induced force is not
universal and is intrinsically related to the physics ruling the
interaction between the elastic medium and the immersed plates
(through the cutoff). Since the short wavelength fluctuations of
the $q$ strings are mainly responsible for the force fluctuations,
the scaling behavior of Eq. (\ref{varianceF}) is expected to hold
for any macroscopic external object with smooth shape at the scale
$a$. The force fluctuations also include a subdominant universal
part $\Delta\!' F$, which can be exactly calculated as
\begin{equation}
\label{varianceF0} \Delta\!' F=(d A_d)^{1/2}
\left(\frac{L}{H}\right)^\frac{d-1}{2}\frac{\kt}{H},
\end{equation}
This cutoff independent part of the dispersion originates from the
$N_H=(L/H)^{d-1}$ strings with $q\lesssim H^{-1}$, as in the
case of the mean Casimir force.

The $\sqrt{N_a}$ factor in Eq.~(\ref{varianceF}) can easily be
understood from the central limit theorem applied to the extensive
system of $2N_a$ non-interacting strings (for $d>1$),
generating a force with mean and standard deviation
of order $\kt/a$. The central limit theorem holds
since for all the strings $\Delta f_{\bf q}(H)/\Delta F\to0$ as
$N_a\to\infty$. As a consequence,
the distribution $\p_{F}$ of the force experienced by
the plate at $x=H$  is a Gaussian:
\begin{equation}\label{pF}
\p_F(F=f)=\frac{1}{\sqrt{2\pi\Delta F}}
\exp{\left[-\frac{\left(f-\langle F\rangle\right)^2}
{2(\Delta F)^2}\right]},
\end{equation}
with its mean and variance being given by Eqs. (\ref{moyenneF})
and (\ref{Fvariance}), respectively. Only the average force felt
by the macroscopic objects is universal, and the dispersion of the
force is essentially controlled by the plates-medium microscopic
interactions.

\textit{Estimation of the Casimir force dispersion}: Assuming that
the instantaneous fluctuation-induced force can be measured with a
perfect precision, the noise over signal ratio is thus given by
Eq.~(\ref{primeresult}), obtained from Eqs. (\ref{moyenneF})
and~(\ref{varianceF}) using $N_{a}=(L/a)^{d-1}$. Let us estimate
it in a typical experimental situation, e.g., a nematic liquid
crystal with a director strongly anchored normal to the plates. In
this case, we have $d=3$ and $a\simeq 1\,\mathrm{nm}$. If the
typical size of the plates is $L\simeq10\,\mu\mathrm{m}$ and
their separation is $H\simeq100\,\mathrm{nm}$, then the amplitude
of the fluctuations is a \textit{hundred times} larger than the
mean force $\langle F\rangle\simeq1\,\mathrm{pN}$. In order to
lower the noise over signal ratio to $1$, the size of the plates
must reach $1\,\mathrm{mm}$!

In all the previous discussions we have implicitly assumed $d>1$.
It is interesting to notice the singularity of the one dimensional
(1D) case. The net force experienced by the plate can no longer be
interpreted as the result of a large number of independent
contributions. A 1D elastic medium is indeed equivalent to the
sole $q=0$ string. Consequently the distribution of forces $F(H)$ and
$F(\infty)$ are directly given by Eq.~(\ref{distributionfq})
taking the limit $q\to 0$. However, even if the distribution is no longer Gaussian,
the scaling of the noise over
signal ratio in Eq. (1) still holds in the limit $d\to1$.

{\it Measurement of fluctuation--induced forces and temporal
resolution}: It has just been shown that the equilibrium
distribution of fluctuation--induced forces may be extremely
broad. Any experimental measurement will however provide a filtered signal which
is averaged over the temporal resolution $\tau$ of the apparatus.
In order to estimate to what extent the
fluctuating nature of fluctuation--induced forces may be
experimentally revealed, a dynamical description is required.
A precise description of the motion of the plates is beyond the
scope of this tentative approach. In a  simple
picture, the measurement apparatus provides a signal
$F_{\tau}(t)=\int_{-\infty}^{t}\chi(t-t')F(t')dt'$. Its response function
$\chi$ has a typical decay time $\tau$
and is causal
and normalized $(\int_{0}^{\infty}\!dt'\,\chi(t')=1)$.
The limit $\tau\to0$ would correspond to a perfect apparatus which provides a
signal completely described by Eq. (\ref{pF}).

The short wavelength excitations of the strings dominate the
fluctuations of the net force as it has been previously shown.
When measuring the force, the apparatus averages over
$N_\tau\simeq\tau/\tau_{a}$ independent processes, in which $\tau_{a}$ is the
microscopic correlation time
of the fluctuations associated to the modes with wavevector
$\simeq a^{-1}$ in the $x$ direction. Consequently, we expect the force
dispersion $\Delta F$ to be lowered by a factor of $1/\sqrt{N_{\tau}}$,
which leads to the result reported above in Eq. (\ref{2emeresult}).

In order to check this analysis, we have studied the simplest case of
a local and dissipative dynamics for the $\phi_{\bf
q}$ fields described by the Langevin equations
\begin{equation}
\gamma\partial_{t}\phi_{\bf q}(x,t)=K \left[ \partial^2_{x} -q^2
\right]
\phi_{\bf
q}(x,t)
+\xi_{\bf
q}(x,t),\label{Langevinq}
\end{equation}
subject to $\phi_{\bf q}(0,t)=\phi_{\bf q}(H,t)=0$. In the above
equation $\xi_{\bf q}$ is a Gaussian white noise with zero mean
and correlations chosen so as to insure thermal equilibrium, i.e.
$\langle\xi_{\bf q}(x,t)\xi_{{\bf q}'}(x',t')\rangle=2\gamma \kt
(2\pi)^{d-1}\delta^{d-1}({\bf q}+{\bf q}') \delta(x-x')
\delta(t-t')$. This set of $N_{a}$ equations correspond to the
Rouse dynamical equations  for the $N_{a}$ pinned elastic
stings undergoing thermal fluctuations~\cite{doiedwards}.
The linearity of the Langevin equations above [Eq.
(\ref{Langevinq})] ensures that the $\phi_{\bf q}$'s are
Gaussian fields, so that it is straightforward to determine the statistical
properties of the measured force.

The finite temporal
resolution $\tau$ does not modify the mean force which
remains given by Eq.~(4), $\langle F_{\tau}\rangle=\langle F \rangle$.
The dispersion of
the measured forces can be expressed as the product of the ideal
expressions calculated previously by an attenuation factor:
\begin{equation}
\Delta F_{\tau}=\Delta F \; \Upsilon(\tau /\tau_a),\label{varianceFtau}
\end{equation}
where for this dynamics $\tau_a$ is given by $\tau_a=\gamma a^2/K$. The
asymptotic scaling behavior of $\Upsilon(\tau/\tau_a)$ does not depend on
the specific form of the response function $\chi$: we have
$\Upsilon(0)=1$ and $\Upsilon(s)\sim 1/\sqrt{s}$ in the limit $s
\gg 1$. This simple model of a diffusive dynamics
thus clearly corroborates the qualitative analysis that lead to Eq.
(\ref{2emeresult}) above.

Let us determine the effect of the finite temporal resolution for the
experimental example quoted above ($L=10\,\mu{\rm m}$,
$H=100\, {\rm nm}$, $a=1\, {\rm nm}$). The order of magnitude of the
relaxation time $\tau_{a}$ in an experiment involving a
nematic liquid crystals is around $0.1\, \mu {\rm s}$ at $T=300 {\rm
K}$. Using a measurement device characterized by $\tau \sim 1\, {\rm
ms}$ such as an optical tweezer \cite{tweezer}, the noise over
signal ratio is lowered from $100$ to $1$.

Generally, Eq.~\ref{2emeresult} quantifies to what extend a slow
measurement device is best suited for experimental observation of
the universal part of fluctuation--induced forces. From the above analysis
it also follows that replacing the fluctuation--induced force between small
objects by the simple
Casimir average when studying their interactions and collective behavior
may not be justified if these
objects are fast movers.

One of us (RG) would like to acknowledge ESPCI for hospitality
during his visit, and support through the Joliot visiting chair.


\begin{references}

\bibitem{Casimir}
H.B.G. Casimir, Proc. K. Ned. Akad. Wet. {\bf 51}, 793 (1948).
%
\bibitem{raminrmp}
M. Kardar and R. Golestanian, Rev. Mod. Phys. {\bf 71}, 1233
(1999), and references therein.
%
\bibitem{goulian}
M. Goulian, R. Bruinsma, and P. Pincus, Europhys. Lett. {\bf 22}, 145
(1993).
%
\bibitem{pauljbf}
P. G. Dommersnes and J.-B. Fournier, Eur. Phys. J. B {\bf 13}, 9 (1999).
%
\bibitem{dietrich}
A. Hanke, F. Schlesener, E. Eisenriegler and S. Dietrich, Phys. Rev. Lett. {\bf
81}, 1885, (1998).
%
\bibitem{aathese}
A. Ajdari, L. Peliti and J. Prost, Phys. Rev. Lett. {\bf 66}, 1481 (1991).
%
\bibitem{ziherl}
P. Ziherl, R. Podgornik, and S. Zumer. Phys.Rev. Lett. {\bf 84}, 1228 (2000).
%
%
\bibitem{kardarli}
H. Li and M. Kardar, Phys. Rev. Lett. {\bf 67}, 3275 (1991);
%
\bibitem{krech}
M. Krech, The Casimir Effect in Critical Systems (World Scientific, Singapore,
1994).
%
%
\bibitem{lamoreaux}
S.K. Lamoreaux, Phys. Rev. Lett. {\bf 78}, 5 (1997).
%
\bibitem{mohideen}
U. Mohideen and A. Roy, Phys. Rev. Lett. {\bf 81}, 4549 (1998).
%
\bibitem{barton}
G. Barton, J. Phys. A. Math. Gen. {\bf 24}, 991 (1991).
%
\bibitem{garcia}
R. Garcia and M. H. W. Chan, Phys. Rev. Lett. {\bf 83}, 1187 (1999). Physica B
{\bf 280} (2000).
%
\bibitem{plasma}
 In the quantum electrodynamic context, where
equation~(\ref{primeresult}) needs to be modified to properly
take into account the temporal dimension, the plasma frequency
of the plates can act as a cut-off~\cite{barton}.
%
\bibitem{law}
A. Mukhopadhyay and B. M. Law, Phys. Rev. Lett. {\bf 83}, 772 (1999).

%
%
\bibitem{deGennes}
P.-G. de Gennes and J. Prost, {\it The Physics of Liquid Crystals}
(Clarendon, Oxford, 1993).
%
\bibitem{landau}
L.D. Landau and E.M. Lifshitz, {\it The Classical Theory of
Fields} 4th edition (Pergamon, Oxford, England, 1975).
%
\bibitem{JZJ}
J. Zinn-Justin, {\it Quantum Field Theory and Critical Phenomena}
2nd edition (Oxford University Press, Oxford, 1993).
%
\bibitem{delta0}
This comes from
$\partial_x\partial_y\mathrm{max}(x,y)|_{x=y}=-2\delta(0)$,
evaluated in Fourier space $-2\int_{-1/a}^{1/a}\!\frac{dk}{2\pi}$,
using the field's cutoff.
%
%
\bibitem{doiedwards}
M. Doi and S.F. Edwards, {\it The Theory of Polymer Dynamics}
(Clarendon, Oxford, 1986).
%
\bibitem{tweezer}
See, e.g.: K. Svoboda and S.M. Block, Annu. Rev. Biophys. Biomol.
Struct. {\bf 23}, 247 (1994).
%
\end{references}
\end{document}